\documentclass[conference,a4paper]{IEEEtran}

\usepackage[dvips]{graphicx, color,epsfig}
\begin{document}

\sloppy

\title{Secure Wireless Communications via Cooperative Transmitting}

\author{
  \IEEEauthorblockN{Toni Draganov Stojanovski}
  \IEEEauthorblockA{University for Information Science and Technology "St. Paul the Apostle"\\
Ohrid, Macedonia\\
    Email: toni\_stojanovski@ieee.org} 
  \and
  \IEEEauthorblockN{Ninoslav Marina}
  \IEEEauthorblockA{Ecole Polytechnique Federale de Lausanne\\
    Lausanne, Switzerland\\
    Email: ninoslav.marina@epfl.ch}
}



\maketitle

\begin{abstract}
Information theoretic secrecy is combined with cryptographic secrecy to create a secret-key exchange protocol for wireless networks.
A network of transmitters, which already have cryptographically secured channels between them, cooperate to exchange a secret key with a new receiver at a random location, in the presence of passive eavesdroppers at unknown locations.
Two spatial point processes: homogeneous Poisson process and independent uniformly distributed points are used for the spatial distributions of transmitters and eavesdroppers.
We analyse the impact of the number of cooperating transmitters and the number of eavesdroppers on the area fraction where secure communication is possible.
Upper bounds on the probability of existence of positive secrecy between the cooperating transmitters and the receiver are derived.
The closeness of the upper bounds to the real value is then estimated by means of numerical simulations.
Simulations also indicate that a deterministic spatial distribution for the transmitters e.g. hexagonal and square lattices, increases the probability of existence of positive secrecy capacity compared to the random spatial distributions.
For the same number of friendly nodes, cooperative transmitting provides a  dramatically larger secrecy region than cooperative jamming and cooperative relaying.
\end{abstract}

\section{Introduction}
Information theoretic secrecy has attracted a significant  interest in recent years due to its possible applications in wireless communications,
and the growing significance of wireless networks.
Wyner \cite{Wyner:1975} first introduced the concept of wiretap channel in 1975.
For discrete memoryless channels, he has determined that a message can be transmitted reliably from a
transmitter to a receiver without revealing any information on the message to the eavesdropper
provided that the transmitter operates at rates smaller than the secrecy capacity.
If the main channel and the wiretap channel are additive white Gaussian noise channels, 
then the secrecy capacity is equal to the difference of the capacities of the two channels as shown by 
Leung-Yan-Cheong and Hellman in \cite{Hellman:1978}.
Csisz{\'a}r and K{\"o}rner \cite{Csiszar-Korner78} extended the previous results to the case of a broadcast channel with confidential
messages.

Secrecy capacity can be improved using cooperation with friendly nodes.
In the {\em cooperative jamming} \cite{Tekin_Yener_2007}, friendly nodes, which are close to the eavesdropper, 
jam the eavesdropper to help increase the achievable secrecy rates for the transmitter
by decreasing the signal-to-noise (SNR) ratio at the eavesdropper.
In the {\em cooperative relaying} \cite{Lai-ElGamal2008, MarinaBH09}, friendly nodes which are closer to the receiver than to the eavesdropper are used as relays.
The relays increase SNR more at the receiver than at the eavesdroppers.

Information theory achieves perfect secrecy as opposed to the computational secrecy provided by cryptographic algorithms.
Here we examine the possibility for mutual applications of cryptographic secrecy and information-theoretic secrecy.
A set of transmitters (e.g. base stations) have already cryptographically secured the communication channels  between them.
When a transmitter wants to communicate securely with a new receiver (e.g. a mobile station),
a pre-secret key message is created by the transmitter, broken into several data blocks, and a separate block is encrypted and sent to each of the other transmitters.
Then each transmitter sends its data block to the receiver.
The transmitters ensure that all the data blocks are received correctly at the receiving node, which is required for the computation of the secret key at the receiver.
The secret key is securely and cooperatively transmitted to the receiver (without being divulged to the eavesdroppers) if the secrecy capacity is positive for the communication channel between at least one transmitter and the receiver.
As the number of transmitters grows, the eavesdroppers are facing a more difficult task of being able to intercept a larger number of transmitters.
Once the secret key is exchanged, the legitimate parties can start communicating at the maximum data rate since their
communication channel is cryptographically protected achieving computational secrecy \cite{Shannon1949}.

Here is the overview of this paper.
In Section~\ref{sec:system_model} we present the system model. 
In Section~\ref{sec-CoopTx} we address the main research questions of this paper: (i) evaluation of the impact of the spatial distribution of transmitters and eavesdroppers on the secrecy region fraction, and derivation of upper bounds for this fraction; 
and (ii) comparison with cooperative relaying and cooperative jamming.
Section~\ref{sect:conclusion} concludes the paper.

\section{Network model} \label{sec:system_model}

We consider two-dimensional wireless networks with the following communication nodes: A network of $L_T$ cooperating transmitters, a single receiver, 
and a network of $L_E$ passive eavesdroppers.
The passive eavesdroppers do not transmit any signal, and try to intercept the information that is transmitted between the pairs of legitimate
nodes, hence reducing the secrecy capability of the network.
Their locations are unknown to the transmitters.
Each transmitter is equipped with only a single omni-directional antenna.

In the sequel, we use the following notation:
\begin{description}
\item [$L(A)$] The area of a region $A\in R^2$;
\item [$L_T$] A random variable which denotes the number of transmitters in a region $A$;
\item [$L_E$] A random variable which denotes the number of eavesdroppers in a region $A$;
\item [$b||c$] A concatenation of two data blocks $b$ and $c$;
\item[$V, V_e$] The additive noise at receiver and eavesdropper, which are independent zero mean Gaussian random variables with variance $\sigma^2$;
\item[$C_{t,r}$] Capacity of the communication channel between transmitter $t$ and receiver $r$;
\item[$C_{s:t,r}$] Secrecy capacity between transmitter $t$ and receiver $r$;
\item[$C_s$] Secrecy capacity between a set of cooperating transmitters and a receiver;
\item[$d_{j,i}$] The distance between nodes $i$ and $j$.
\end{description}

We use the additive white Gaussian noise model.
Then, the received signal at the receiver $r$ from the transmitter $t$ is 
\begin{eqnarray*}
Y = d_{t,r}^{-\beta/2}X+V.
\end{eqnarray*}
where $X$ is the transmitted signal from the transmitter $t$, and $\beta$ is the path-loss coefficient \cite{Rappaport:book}.
The received signal at the eavesdropper $e$ from the transmitter $t$ equals
\begin{eqnarray*}
Z_e = d_{t,e}^{-\beta/2}X+V_e.
\end{eqnarray*}

The point to point capacities between transmitter $t$ and receiver $r$, and between transmitter $t$ and  eavesdropper $e$ are given by \cite{Hellman:1978}
\begin{eqnarray}
C_{t,r}&=&\frac{1}{2}\log_2\left(1+\frac{P_t d_{t,r}^{-\beta}}{\sigma^2}\right)\nonumber\\
C_{t,e}&=&\frac{1}{2}\log_2\left(1+\frac{P_t d_{t,e}^{-\beta}}{\sigma^2}\right)\label{eq_SecRegNonCoop}
\end{eqnarray}
where $P_t$ is the transmitter's power.
If the point to point capacity between the transmitter and the eavesdropper $C_{t,e}$ is
larger than the capacity of the channel between the two communicating nodes $C_{t,r}$, then $C_{s:t,r}=0$.
Otherwise, $C_{s:t,r}>0$ \cite{Hellman:1978}:
\begin{eqnarray*}
C_{s:t,r}=\max\{C_{t,r}-C_{t,e}, 0\}
\label{eq_cs}
\end{eqnarray*}
From Eq.~(\ref{eq_SecRegNonCoop}) it follows that $C_{s:t,r}>0$ if the receiver $r$ is closer to the transmitter than the eavesdropper, that is, $d_{t,r}<d_{t,e}$.
The disk $D_s \subset R^2$ with center at the transmitter and radius equal to the distance between the transmitter and the nearest eavesdropper is called {\em secrecy disk} of the transmitter.
If a receiver is inside the secrecy disk, then the secrecy capacity between the transmitter and the receiver is positive.

Receivers which are outside the secrecy disk for a given transmitter can not communicate securely with that transmitter.
In the next section we explain a type of cooperation for a set of friendly transmitter that combines their secrecy disks 
and thus allows them to communicate secretly with receivers positioned in a larger region.

\section{Cooperative transmitting}\label{sec-CoopTx}
The set of transmitters have already established a cryptographic secret key, and they can cryptographically protect their mutual communication channels.
Let assume that transmitter $t_i$ and a new communicating node/receiver $r$ want to communicate.
$t_i$ generates a pre-secret key message $B$ with arbitrary length, which it then divides into $L_T$ blocks $b_1, b_2, \dots, b_{L_T}$.
Each block is sent to a a different transmitter via a cryptographically secured channel.
Then each transmitter $t_i$ sends its block $b_i$ to the receiver.
The intended receiver correctly receives all blocks $b_1, b_2, \dots, b_{L_T}$, and restores the pre-secret key message $B=b_1 || b_2 || \dots || b_{L_T}$. 
Both $t_i$ and $r$ use a cryptographic hash function $H$ to calculate the mutual secret key $K=H(B)$, which is then used to cryptographically protect their mutual communication.
The eavesdroppers have to be able to intercept the transmission from all $L_T$ transmitters.
If at least one data block out of $L_T$ data blocks is not intercepted, then the secret key $K$ can not be computed at the eavesdropper. 
We call this strategy for cooperation - {\em cooperative transmitting}.
Using cooperative transmitting a transmitter can exchange a secret key with a receiver if the receiver is inside any of the secrecy disks for all $L_T$ transmitters.
The impact of cooperative transmitting is quantitatively measured through the fraction $F_s(A)$ of a region $A$ covered by the  union of secrecy disks.
In other words, fraction $F_s(A)$ is equal to the probability of securely exchanging a secret-key with a receiver that is randomly positioned inside the region $A$
\begin{eqnarray*}
F_s(A)\equiv \mathbf{P}\{C_s>0\}
\end{eqnarray*}.
The coverage problem by secrecy disks was studied by Sarkar and Haenggi \cite{SarkarHaenggi-2010}. 
They studied the covered volume fraction and the asymptotic conditions for complete coverage in one and two dimensions.

\begin{figure}[htbp]
\centering
\includegraphics[width=80mm]{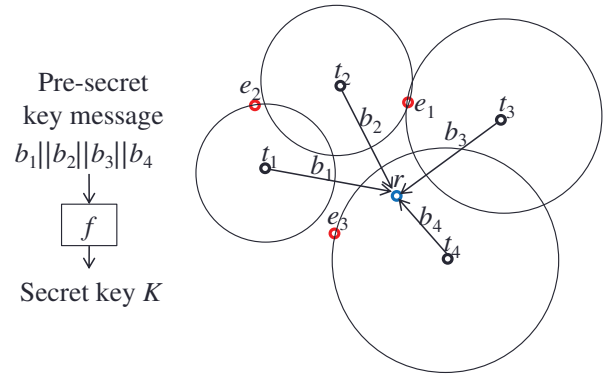}
\caption{Sample network with four transmitters, one new receiver, and three eavesdroppers.}
\label{fig-SampleNetwork}
\end{figure}

Figure~\ref{fig-SampleNetwork} illustrates the concept of cooperative transmitting on a sample network.
Receiver $r$ is inside the secrecy disk of transmitter $t_4$, and therefore block $b_4$ can not be intercepted by any of the eavesdroppers.

In the remainder of this paper, we analyse the dependence of $\mathbf{P}\{C_s>0\}$ on the spatial distributions of transmitters and eavesdroppers.
We analyse both random and deterministic models for the spatial distribution of transmitters and eavesdroppers.
Two simple models for random spatial processes for the transmitters and the eavesdroppers will be used.
The first model is homogeneous Poisson process on the plane characterised by the mean number of points $\lambda$ in a unit area, called also {\em rate} or {\em density} of the Poisson Process.
The number of points $l$ inside a region $A$ follows the Poisson probability distribution law with parameter $\lambda L(A)$
\begin{equation}\label{eq-Poisson}
P_{L}(l) = \frac{(\lambda L(A))^{l}}{l!}e^{-\lambda L(A)}
\end{equation}
In the second model a fixed number of points are independently and uniformly distributed (IUD) in a certain region of the plane, characterised by s single parameter - the fixed number of points. 
These models are widely used in the literature on information theoretic secrecy \cite{SarkarHaenggi-2010,SarkarHaenggi2011,Pinto2009}, the reason being twofold.
They provide a good first-order approximation for the spatial distribution of communication nodes in real networks.
Second, simplicity of the homogeneous Poisson process and IUD process allows for an analytical analysis of information security-related metrics e.g. fraction $F_s(A)$.
For the spatial distribution of the transmitters we will also investigate two deterministic models: hexagonal lattice and square lattice.

\subsection{IUD transmitters and IUD eavesdroppers}\label{sec-IUDTxIUDE}
In the first case, the position of the transmitters in a region $A\in R^2$ obeys a IUD process with parameter $n_T$. Similarly, a fixed number of eavesdroppers $n_E$ are positioned according to an IUD process in the same region $A$.
If $n_T=1$, then $C_s>0$ if the receiver is inside the secrecy disk of the transmitter, that is, it is closer to the transmitter than any of the $n_E$ eavesdroppers:
\begin{equation}\label{nteq1}
\mathbf{P}\{C_s>0\} = \frac{1}{1+n_E}
\end{equation}

For $n_T>1$, we establish an upper bound for $\mathbf{P}\{C_s>0\}$ as follows.
For $n_T=2$, the secrecy region of the two transmitters is union of their secrecy disks:
\begin{eqnarray*}
\mathbf{P}\{C_s>0\}=1-\mathbf{P}\{C_s<0\}\leq\\
1-\mathbf{P}\{C_{s:1,r}<0\}\mathbf{P}\{C_{s:2,r}<0\}\leq 1-\left(\frac{n_E}{1+n_E}\right)^{2}
\end{eqnarray*}
where the overlapping area of the two secrecy disks is neglected in the upper bound.
One can generalise for $n_T>1$ thus obtaining
\begin{equation}\label{eq-c1e1}
\mathbf{P}\{C_s>0\}\leq 1-\left(\frac{n_E}{1+n_E}\right)^{n_T}
\end{equation}

Next we consider the case when both $n_T$ and $n_E$ grow infinitely, while their ratio remains constant $k=\frac{n_T}{n_E}$. 
This is a good first order approximation when the area of region $A$ grows infinitely and the densities of transmitters and eavesdroppers remain constant.
Then
\begin{equation}\label{eq-c1e2}
\lim_{n_E\to\infty}\mathbf{P}\{C_s>0\}\leq \lim_{n_E\to\infty}1-\left(\frac{n_E}{1+n_E}\right)^{kn_E} = 1-e^{-k}
\end{equation}

\begin{figure}[htbp]
\centering
\includegraphics[width=90mm]{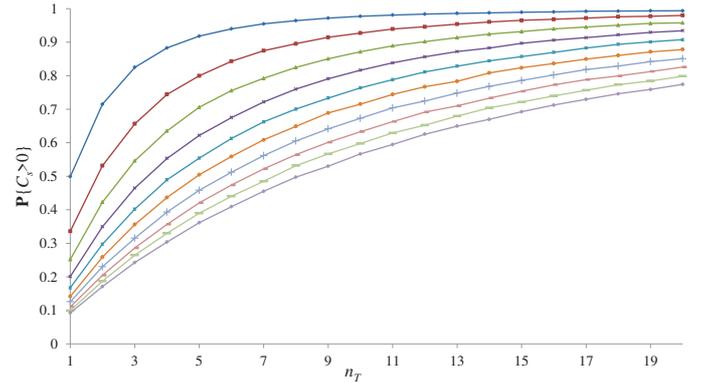}
\caption{Dependence of $\mathbf{P}\{C_s>0\}$ on the number of transmitters $n_T$ with $n_E$ as the curves' parameter.
The lowest curve corresponds to $n_E=10$, and the highest curve is for $n_E=1$.}
\label{fig-CsvsnTIUDIUD}
\end{figure}

In order to evaluate the closeness of the upper bounds (\ref{eq-c1e1}) and (\ref{eq-c1e2}) to the real value,
we have numerically estimated the value for $\mathbf{P}\{C_s>0\}$.
Figure~\ref{fig-CsvsnTIUDIUD} depicts the dependence of $\mathbf{P}\{C_s>0\}$ on $n_T$ and $n_E$ as obtained from the numerical simulations.
Each point on the curves is obtained from 100,000 network simulations.

\begin{figure}[htbp]
\centering
\includegraphics[width=90mm]{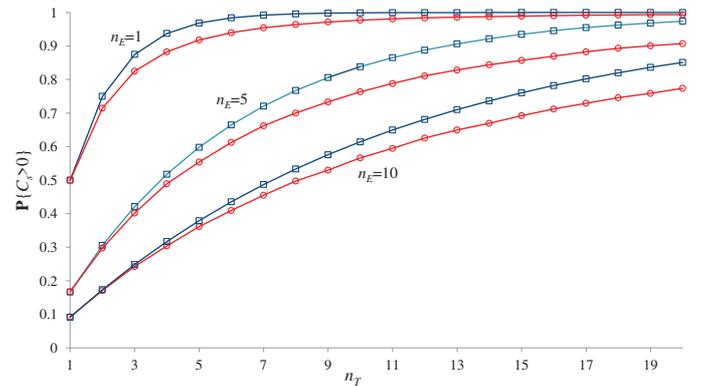}
\caption{Closeness between the real value for $\mathbf{P}\{C_s>0\}$  (circles) and the upper bound (squares) given by Eq.~(\ref{eq-c1e1}).}
\label{fig-IUDIUD_UpperBound}
\end{figure}

Figure~\ref{fig-IUDIUD_UpperBound} shows the closeness between upper bound given by Eq.~(\ref{eq-c1e1}) and the real values for $\mathbf{P}\{C_s>0\}$, which are estimated through numerical simulations.
Relative gap between the upper bound and the real values grows for larger $n_T$ due to the increasing number of overlapping secrecy disks.

\begin{figure}[htbp]
\centering
\includegraphics[width=90mm]{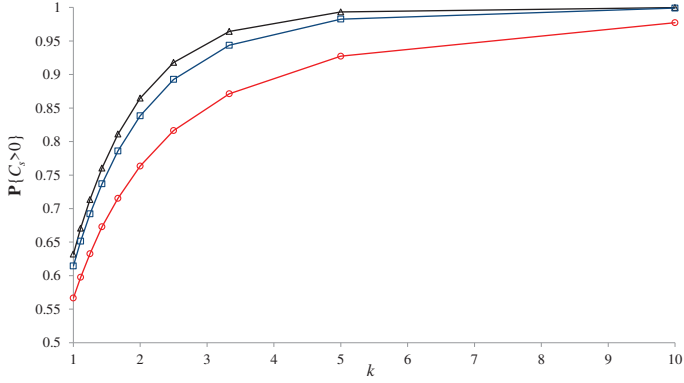}
\caption{Closeness between the real value for $\mathbf{P}\{C_s>0\}$ (circles), and  the upper bounds given by Eqs.~(\ref{eq-c1e1}) (squares) and (\ref{eq-c1e2}) (triangles). $n_T=10$ and $n_E=1,2,\dots, 10$. }
\label{fig-IUDIUD_ThreeCurves}
\end{figure}

Figure~\ref{fig-IUDIUD_ThreeCurves} shows the closeness between the upper bounds (\ref{eq-c1e1}) and (\ref{eq-c1e2}), and the numerically estimated values for $\mathbf{P}\{C_s>0\}$.
Relative  gap between the upper bounds and the real values gets smaller for smaller $k$ (larger $n_E$) since the secrecy disks as well as their overlaps become smaller in size.

\subsection{Poisson transmitters and IUD eavesdroppers}\label{sec-PoissonTxIUDE}
Next we consider the case where transmitters are positioned according to a Poisson spatial process with rate $\lambda_T$.
Without lose of generality of the results, we assume that $L(A) =1$ and thus the average number of transmitters in the region $A$ is $\lambda L(A)=\lambda$.
Eavesdroppers' positions obey an IUD process and the number of eavesdroppers in the region $A$ is $n_E$.

If the number of transmitters $L_T$ is 1, then Eq.~(\ref{nteq1}) holds.
For $L_T>1$, the upper bound given by Eq.~(\ref{eq-c1e1}) is valid.
Then an upper bound for $\mathbf{P}\{C_s>0\}$ can be derived as an average value of functions (\ref{nteq1}) and (\ref{eq-c1e1}) for the random variable $L_T$:
\begin{eqnarray}
& &\mathbf{P}\{C_s>0\}=E\left[\mathbf{P}\{C_s>0|L_T\} \right]\nonumber\\ 
& \leq&\frac{1}{1+n_E}\lambda_T e^{-\lambda_T} + \sum_{l_T=2}^{\infty}\left(1-\left(\frac{n_E}{1+n_E}\right)^{l_T}\right)\frac{\lambda_T^{l_T}}{l_T!}e^{-\lambda_T}\nonumber\\
& =&1-e^{\frac{-\lambda_T}{1+n_E}}\label{eq-UB_PoissonIUD}
\end{eqnarray}

\begin{figure}[htbp]
\centering
\includegraphics[width=90mm]{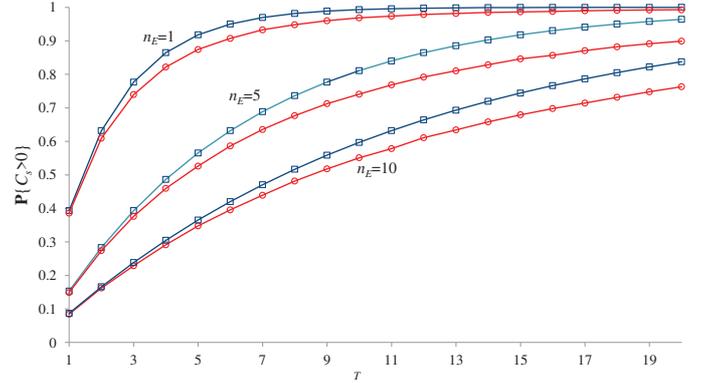}
\caption{Closeness between the real value for $\mathbf{P}\{C_s>0\}$, and  the upper bound given by Eq.~(\ref{eq-UB_PoissonIUD}).}
\label{fig-UBCloseness_PoissonIUD}
\end{figure}

Figure~\ref{fig-UBCloseness_PoissonIUD} shows the closeness between the upper bound (\ref{eq-UB_PoissonIUD}), and the numerically calculated values for $\mathbf{P}\{C_s>0\}$.
Similar to Fig.~\ref{fig-IUDIUD_UpperBound}, accuracy of the upper bound decreases for larger $\lambda_T$ as a consequence of the increasing number of intersecting secrecy disks.
Numerical simulation of a Poisson spatial process was done according to \cite{Moltchanov-2012}.
In order to generate a Poisson process with rate $\lambda$ in a region $A$, we first randomly select a value $l$ for a Poisson variable with mean $\lambda L(A)$,
and then we randomly position $l$ IUD points in $A$.
Observed dependence of $\mathbf{P}\{C_s>0\}$ on $\lambda_T$ and $n_E$ was similar to the one depicted in Fig~\ref{fig-CsvsnTIUDIUD}.

\addtolength{\textheight}{-3.5cm}

\subsection{IUD transmitters and Poisson eavesdroppers}\label{sec-IUDTxPoissonE}
A fixed number of transmitters $n_T$ are positioned at IUD points in a region $A\in R^2$.
Positions of eavesdroppers follow a Poisson spatial process with average rate $\lambda_E$. 
For sake of simplicity we again assume that $l(A) = 1$
Then the number of eavesdroppers in $A$ is a Poisson random variable $L_E$ with average value $\lambda_E$.
Its probability distribution function is given by Eq.~(\ref{eq-Poisson}) where $\lambda=\lambda_E$.
For $n_T=1$, the secrecy region fraction is given by
\begin{eqnarray*}
\mathbf{P}\{C_s>0\} = E\left[\frac{1}{1+L_E}\right] = \frac{1}{\lambda_E}(1-e^{-\lambda_E})
\end{eqnarray*}

\begin{figure}[htbp]
\centering
\includegraphics[width=90mm]{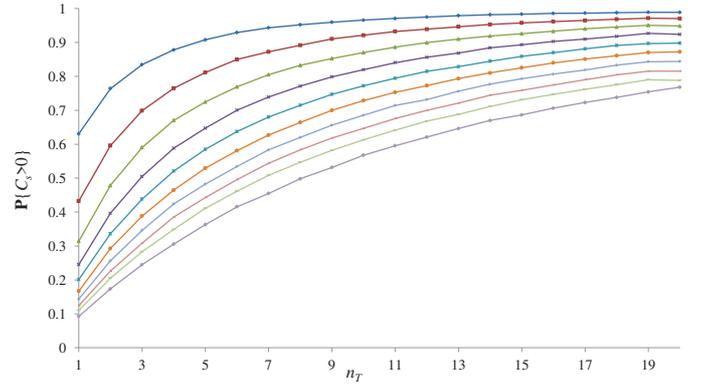}
\caption{Dependence of $\mathbf{P}\{C_s>0\}$ on the number of transmitters $n_T$ with $\lambda_E$ as the curves' parameter.
The lowest curve corresponds to $\lambda_E=10$, and the highest curve is for $\lambda_E=1$.}
\label{fig-CsvsnTIUDPoisson}
\end{figure}

For $n_T>1$ we ran numerical simulations, and the results are given in Fig.~\ref{fig-CsvsnTIUDPoisson}. Note the similarity with Fig.~\ref{fig-CsvsnTIUDIUD}.

For sake of completeness, we have also numerically analysed the case when a homogeneous Poisson process in a region $A\in R^2$ 
is assumed for both transmitters and eavesdroppers.
Again we have obtained very similar results to the previously analysed three combinations of IUD and Poisson spatial processes for transmitters and eavesdroppers.
Following slight differences were observed.
IUD spatial process for the transmitters gives slightly higher values for $\mathbf{P}\{C_s>0\}$ than the Poisson spatial processes.
On the contrary, the Poisson spatial process for the eavesdroppers gives slightly higher values for $\mathbf{P}\{C_s>0\}$ than the IUD spatial processes.

\subsection{Transmitters in deterministic lattice and UID eavesdroppers}\label{sec-GridTxPoissonE}
\begin{figure}[htbp]
\centering
\includegraphics[width=80mm]{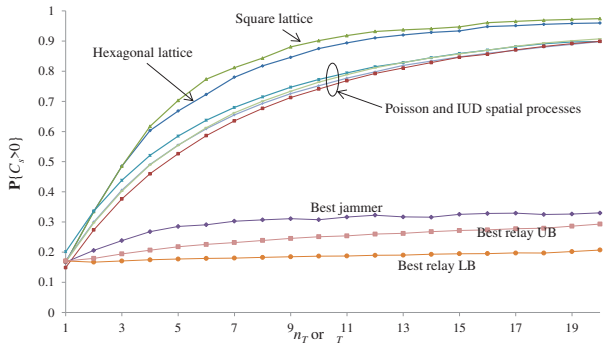}
\caption{Comparison of stochastic and deterministic positioning of eavesdroppers. Top two curves are for hexagonal and square lattice (IUD eavesdroppers with $n_E=5$), while middle four curves are for Poisson and IUD spatial processes for the transmitters' and eavesdroppers'  positions (with $\lambda_E=5$ or $n_E=5$). Bottom three curves are for cooperative jamming and cooperative relaying.}
\label{fig-Comparison}
\end{figure}

Next we analysed the case when the transmitters are positioned on a deterministic lattice, and the eavesdroppers obey a UID process.
By means of numerical simulations we examined a square lattice and a hexagonal lattice.
We observed similar shapes to the curves shown in Fig.~\ref{fig-CsvsnTIUDIUD} and Fig.~\ref{fig-CsvsnTIUDPoisson} for stochastic spatial processes for the transmitters.
$\mathbf{P}\{C_s>0\}$ is higher for a deterministic lattice compared to a stochastic spatial processes for the transmitters (see Fig.~\ref{fig-Comparison})
due to the lower variations in the overlap between the secrecy disks of individual transmitters.
For a stochastic spatial process, there are areas which can be covered by multiple overlapping secrecy disks of nearby transmitters.
At the same time in the regions with sparse transmitters, it is more probable to find subregions not covered by any secrecy disk.

\subsection{Comparison with cooperative jamming and cooperative relaying}\label{sec-comparison}
In this section we compare cooperative transmitting with two other strategies for cooperation in wireless networks.
Cooperative relaying and cooperative jamming increase the secrecy capacity by means of widening the gap between the SNR at the legitimate receiver and the SNR at the eavesdroppers.
In the “single hop cooperation with the best relay” \cite{MarinaStojanovskiPoor-ITA2012, MarinaStojanovskiPoor-2012} only the strongest relay is selected from the set of UID randomly positioned relays, which is the relay node which most improves the secrecy capacity. 
In the "single hop cooperation with the best jammer" \cite{MarinaStojanovskiPoor-ITA2012, MarinaStojanovskiPoor-2012} a single node from the set of friendly nodes is selected to act as a jammer.
Cooperative jamming aims to reduce the SNR at the legitimate receiver, but at the same time it reduces the SNR even more at the eavesdroppers.
On the contrary, the best relay increases the secrecy capacity by increasing SNR at the legitimate receiver more than it increases SNR at the eavesdroppers. 

We use the value for $\mathbf{P}\{C_s>0\}$ as a quantitative measure of the positive impact of the different strategies for cooperation.
Figure~\ref{fig-Comparison} shows that cooperative transmitting offers dramatic improvement in the secrecy region's size over cooperative jamming and cooperative relaying.


%
%
\section{Conclusion} \label{sect:conclusion}
In this work we propose to combine information theoretic secrecy with cryptographic secrecy to increase the secrecy region, and provide a novel solution to the key-exchange problem. 
Cooperative transmitting can significantly improve information-theoretic secrecy in wireless networks.
The type of cooperation is quite important for the resulting secrecy region.
For the same number of friendly nodes, cooperative transmitting provides a  larger coverage area than cooperative jamming and cooperative relaying.

\bibliographystyle{IEEEtran}
\bibliography{ref_isit11c}

\end{document}